\documentclass{ws-procs961x669}            

\newcommand{\as}{\alpha_\mathrm{s}}

\begin{document}
\title{Hadron structure and parton branching beyond\\ collinear approximations\footnote{Contribution 
at the Workshop ``Probing Nucleons and Nuclei in High Energy Collisions" (INT-18-3), 
Institute for Nuclear Theory, University of Washington, Seattle (October 2018)} \\
}

\author{F Hautmann}

\address{University of Oxford / University of Antwerp\\
E-mail: francesco.hautmann@physics.ox.ac.uk}

\begin{abstract}
We briefly illustrate recent developments in the parton branching formulation of 
TMD evolution and their impact on precision measurements in high-energy hadronic collisions.  
\end{abstract}


\bodymatter

\vskip 0.6 cm 

The impact of hadron structure on precision studies of fundamental interactions and 
searches for new physics 
plays an essential role at high-energy colliders of the present and next generation. The 
QCD theoretical framework based on 
collinear parton distribution functions (PDFs) and parton showers, in particular,  is 
extremely successful in describing a wealth of collider data.~\cite{Kovarik:2019xvh}  

However, PDFs are the result of a 
strong reduction of information and tell us only about the longitudinal momentum of partons in a fast moving hadron. 
This restriction is lifted in  the transverse momentum dependent (TMD) parton distribution functions ---  more general distributions which provide 
``3-dimensional imaging" of hadron structure.~\cite{Angeles-Martinez:2015sea}  Such distributions are needed to obtain  QCD factorization 
formulas for  collider  observables in  ``extreme"  kinematic regions characterized by multiple momentum scales. These  will be  relevant  
both for experiments at the high-energy frontier and for exploring the region of   the  highest masses  accessible  at the high-luminosity frontier. 

A large body of knowledge has been built  about 
collinear PDFs  over the last three decades  from the analysis of  high-energy experimental data in hadronic collisions, 
greatly aided by the development of realistic  Monte Carlo  (MC) event 
simulations~\cite{Bengtsson:1987kr,Marchesini:1987cf}  for the parton cascades associated with  PDF evolution. 
TMDs, on the other hand,  are much less known.  Hadronic 3D imaging, with  its 
implications for high-energy physics, will constitute the subject of 
intensive studies in the forthcoming decade.    
 The construction of MC   event 
 generators incorporating 
TMDs and 3D hadron structure effects~\cite{Hautmann:2019rvr}  
is 
thus 
a central objective of 
physics programs for future hadron colliders (HL-LHC, LHeC, EIC, FCC).

Steps toward TMD MCs have recently been  taken in the works~\cite{Martinez:2019mwt,Martinez:2018jxt,Hautmann:2017fcj,Hautmann:2017xtx}, in which a parton branching 
formalism is proposed for TMD evolution, and applications to deep inelastic scattering (DIS) and Drell-Yan (DY) processes are presented. 
In the following we give a brief account of these studies.

The parton branching (PB) approach gives TMD evolution equations of the schematic form~\cite{Hautmann:2017fcj,Hautmann:2017xtx}  
\begin{eqnarray}
\label{eq:tmdevol}
 {A}_a\left( x, {\bm k}, \mu^2\right) &=& 
 \Delta_a\left(\mu^2, \mu_0^{2}\right)
 {A}_a\left( x, {\bm k}, \mu_0^2\right)
  \\ 
 &+& 
 \sum_b\int \frac{\textrm{d}^2{\boldsymbol \mu}^{\prime}}{\pi {\mu}^{\prime 2}} 
   \int  \textrm{d}z 
\   {\cal K}_{a b}    \left(  x, {\bm k}, \mu^2 ;   z , z_M ,  {\mu}^{\prime 2}     \right) 
{A}_b\left( x / z ,  {\bm k} + a(z){\boldsymbol \mu}^\prime, \mu^{\prime 2}\right) \; , 
\nonumber   
\end{eqnarray} 
where $
A_a\left( x, {\bm k}, \mu^2\right)$ is the 
TMD distribution of flavor $a$, carrying the longitudinal momentum  fraction $x$ of the hadron's momentum and  transverse momentum ${\bm k}$ 
at the evolution scale $\mu$;  $z$ and ${\boldsymbol \mu}^\prime$ are the branching variables, with $z$ being the longitudinal momentum transfer  
at the  branching, and  $ \mu^\prime = \sqrt{ {\boldsymbol \mu}^{\prime 2}}$ the momentum scale at which the branching occurs; 
$z_M$ is the soft-gluon resolution scale;  the function $a(z)$ specifies the ordering condition  in the branching; 
 $ {\cal K}_{a b} $    are evolution kernels, computable in terms of  
 Sudakov form factors,   real-emission splitting functions and phase-space constraints.    The initial  
evolution scale is denoted by $\mu_0$;      the   distribution  $ {A}_a\left( x, {\bm k}, \mu_0^2\right)$ at scale $\mu_0$ in the first term on the right hand side of 
Eq.~(\ref{eq:tmdevol})   is the intrinsic   $k_T$  distribution. 

The PB evolution equations  are designed to be applicable over a wide kinematic range from low to high energies and implementable in 
MC generators.  
By taking  the ordering function  $a(z)$, soft-gluon resolution scale $z_M$ and strong coupling $\as$  of  the form prescribed 
by  angular ordering~\cite{Catani:1990rr,antw1908},  
  Eq.~(\ref{eq:tmdevol})   gives, once it is integrated over transverse momenta,  
  the CMW coherent-branching equation.~\cite{Marchesini:1987cf,Catani:1990rr}
On the other hand, for  soft-gluon resolution  
$z_M \to 1$ and strong coupling $\as \to \as (\mu^{\prime 2})$,   integrating  Eq.~(\ref{eq:tmdevol}) over transverse momenta 
gives    collinear PDFs satisfying  
DGLAP evolution equations.~\cite{Gribov:1972ri,Altarelli:1977zs,Dokshitzer:1977sg}
The convergence to DGLAP 
 at leading order (LO) and  next-to-leading order (NLO)  has been 
verified numerically in~\cite{Hautmann:2017xtx} against the evolution program~\cite{Botje:2010ay} 
  at the 
   level of better   than 1\% over a range of five orders of magnitude both in $x$ and in $\mu$.
Besides the collinear limits, Eq.~(\ref{eq:tmdevol}) can be used at unintegrated level for event simulation 
of TMD physics effects.~\cite{jcc-book}

\begin{figure}[htb]
\begin{center} 
\includegraphics[width=0.405\textwidth]{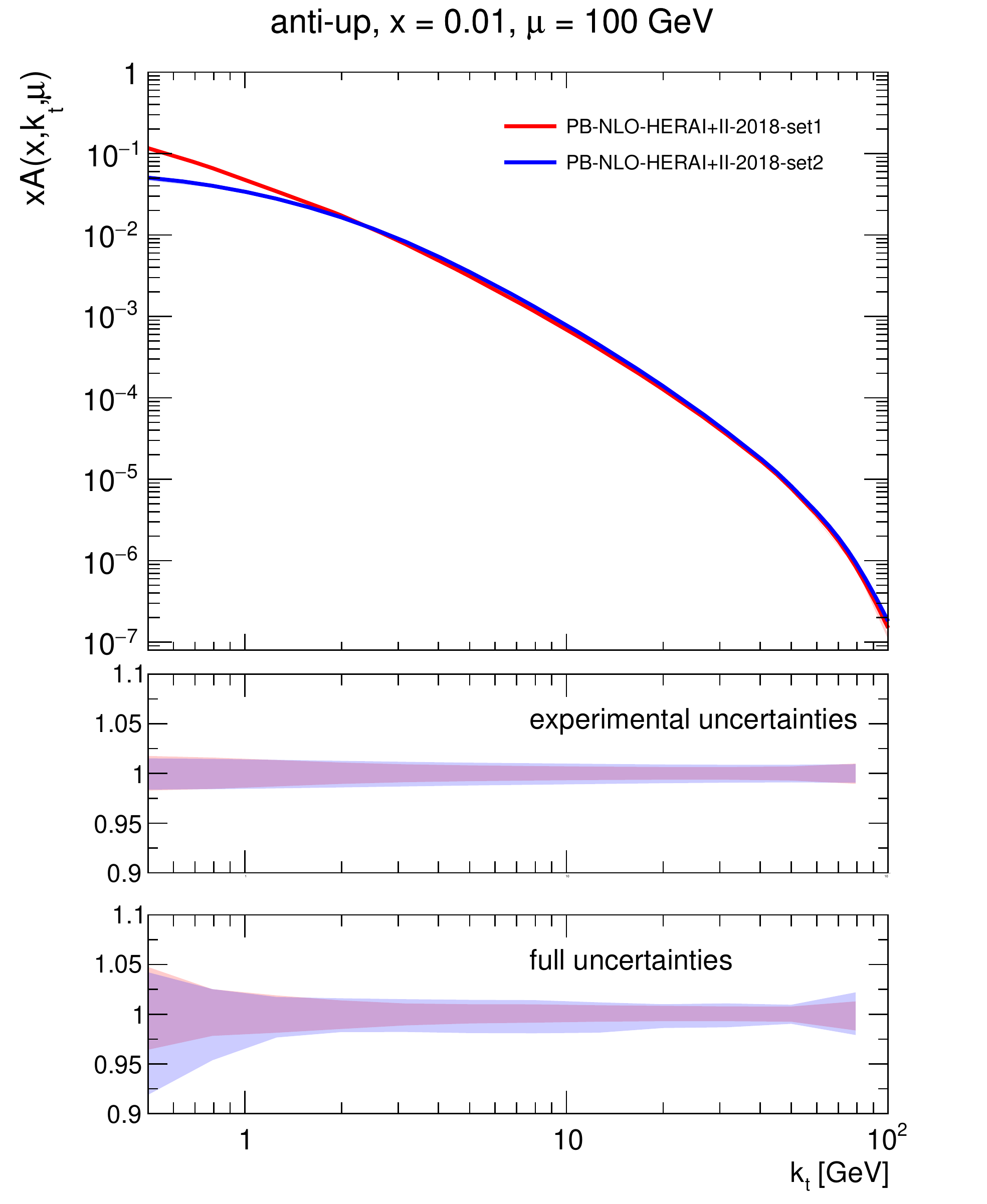}
\includegraphics[width=0.405\textwidth]{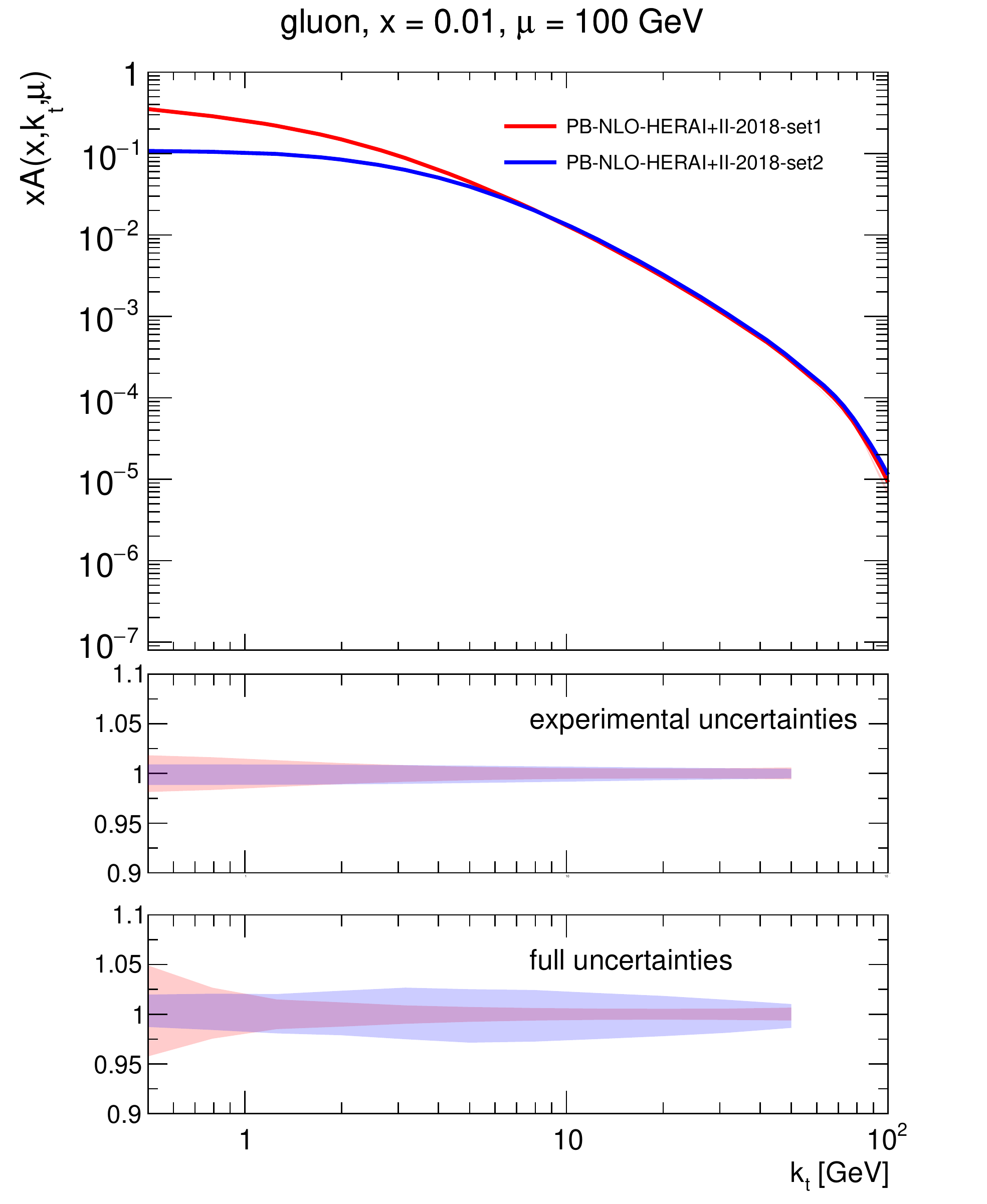}
  \caption{\small TMD   $\bar u$ and gluon  distributions as a function of $k_T$ for  $\mu=100$~GeV at $x=0.01$~\protect\cite{Martinez:2018jxt}. In the lower panels 
 the relative uncertainties are shown,  coming  from experimental uncertainties and the total of experimental and model uncertainties.
  }
\label{TMD_pdfs3}
\end{center}
\end{figure}

\begin{figure}[htb]
\begin{center} 
\includegraphics[width=0.405\textwidth]{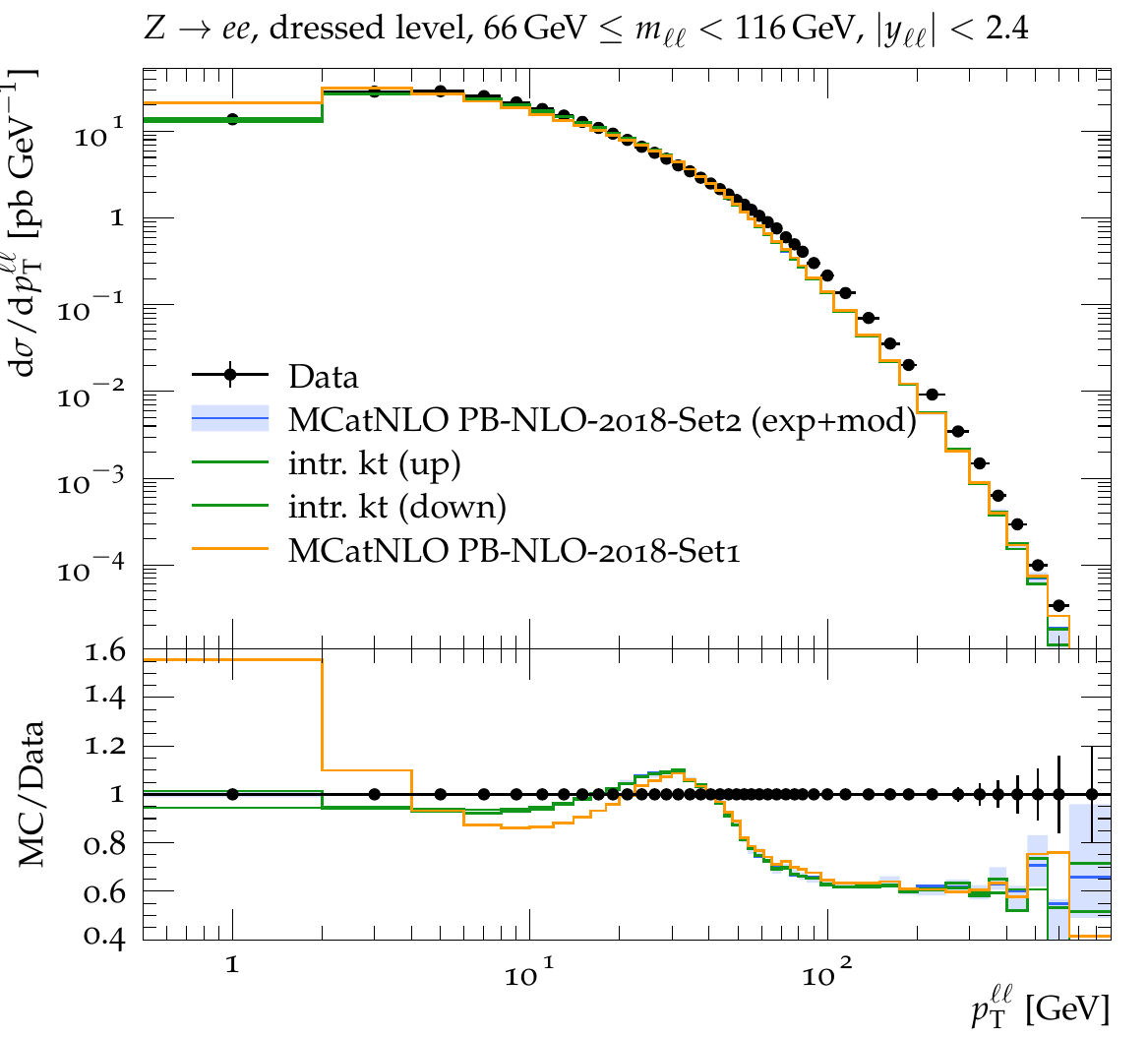} 
\includegraphics[width=0.405\textwidth]{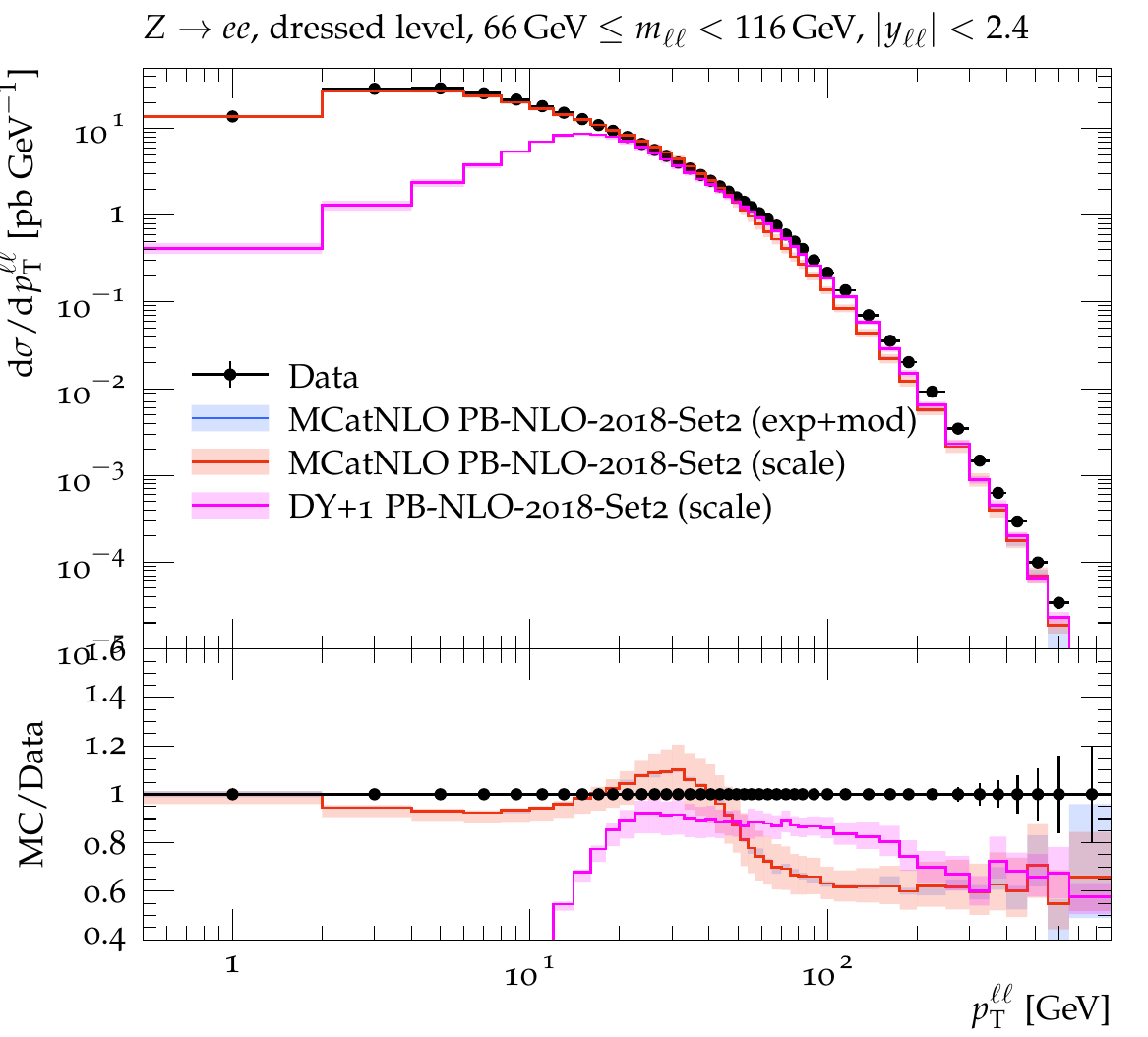} 
\caption{\small Transverse momentum $p_T$ spectrum of Z -bosons as measured by \protect\cite{Aad:2015auj} at $\sqrt{s}=8 $ TeV compared to the 
prediction~\protect\cite{Martinez:2019mwt}  using aMC@NLO and NLO PB -TMD. 
Left: uncertainties from the PB -TMD and 
 from changing the width of the intrinsic gaussian distribution by a factor of two. Right: with uncertainties from the TMDs  and scale variation combined. 
  }
\label{Zpt-TMD_uncertainty} 
\end{center}
\end{figure}

In Figs.~\ref{TMD_pdfs3} and \ref{Zpt-TMD_uncertainty} we give examples of PB-TMD applications to DIS and DY processes. 
Fig.~\ref{TMD_pdfs3}~\cite{Martinez:2018jxt}  shows results for TMDs from PB fits at NLO 
 to the HERA high-precision inclusive 
DIS data~\cite{Abramowicz:2015mha},  performed 
using the   fitting platform   \verb+xFitter+~\cite{Alekhin:2014irh} and the numerical techniques  
developed in~\cite{Hautmann:2014uua} to treat the transverse momentum dependence in the fitting procedure. 
 In~\cite{Martinez:2018jxt}    two fitted TMD sets are presented,  differing  by the 
treatment of the momentum scale in the  coupling $\as$, so that one can 
compare the effects of $\as$ evaluated at the transverse momentum scale  
prescribed by the angular-ordered branching~\cite{Catani:1990rr,Martinez:2018jxt} with $\as$ evaluated at the evolution scale.  
The TMDs are extracted including a determination of experimental and theoretical uncertainties. The lower panels in 
Fig.~\ref{TMD_pdfs3}    show these uncertainties for $\bar u$ and gluon 
 distributions, at fixed values of $x$ and $\mu$, as a function of transverse momentum. 

The  $k_T$ dependence in Fig.~\ref{TMD_pdfs3} results from  intrinsic transverse momentum and  evolution.   
The intrinsic  $k_T$ in  Fig.~\ref{TMD_pdfs3} is taken for simplicity to be 
described by a gaussian at $\mu_0 \sim {\cal O}$ (1 GeV)  
with (flavor-independent and $x$-independent) width $\sigma = k_0 / \sqrt{2}$,  $k_0 = 0.5$ GeV. 
This is to be compared with higher values of intrinsic $k_T \sim$ 2  GeV obtained  from tuning in shower MC event  
generators (see e.g.~\cite{Khachatryan:2015pea}).

In Fig.~\ref{Zpt-TMD_uncertainty}~\cite{Martinez:2019mwt}   the PB  TMDs are combined with the 
NLO calculation of 
 DY  $Z$-boson production to determine predictions  for  the lepton-pair transverse momentum $p_T$  spectrum. 
 These  are  compared   with 
  LHC measurements.~\cite{Aad:2015auj}  
The computation in Fig.~\ref{Zpt-TMD_uncertainty} 
 requires addressing  issues of matching~\cite{jcc-fh-jhep}  analogous to those that arise in the case of parton showers. The 
matching is accomplished in the aMC@NLO framework.~\cite{alwall14}       
The calculations are performed via  
 {\sc Cascade}~\cite{Jung:2010si}   to read LHE~\cite{Alwall:2006yp}  files, perform  TMD evolution,~\cite{Hautmann:2017fcj} produce output files,  
 and {\sc Rivet}~\cite{Buckley:2010ar} to analyze the outputs.

The behaviors in the DY spectrum in Fig.~\ref{Zpt-TMD_uncertainty}  
 can be understood in terms of the  $k_T$  distributions in  Fig.~\ref{TMD_pdfs3}. 
The uncertainties on the DY predictions 
come from TMD uncertainties and scale variations, with the latter dominating the overall uncertainty. 
We see from the left panel in Fig.~\ref{Zpt-TMD_uncertainty}  that the spectrum at low $p_T$ is sensitive to the 
angular ordering effects embodied in the  different treatment of $\as$ in the  PB Set 1 and Set 2. 
The bump in the $p_T$ distribution for intermediate $p_T$ is an effect of  the matching and choice of 
the matching scale~\cite{Martinez:2019mwt,alwall14}   --- a similar effect is seen when using parton showers instead of PB  TMD. 
The deviation in  the spectrum at higher $p_T$ is due to including only  ${\cal O} (\as) $ corrections but missing higher orders. We see from 
 the right panel of Fig.~\ref{Zpt-TMD_uncertainty}  that the contribution from DY + 1 jet at NLO plays an important role at 
 larger $p_T$.  The  merging of higher jet multiplicities~\cite{bermudeztalk} in the PB TMD framework is one  of the 
  ongoing  developments  needed for MC event generators including 3D hadron  structure effects.

\vspace*{0.3cm}  
\noindent {\bf Acknowledgments}. 
I thank  the INT staff and  workshop organizers   for the invitation and hospitality.

\end{document}